\documentclass[aps,prx,twocolumn,superscriptaddress,amsmath,amssymb]{revtex4}

\usepackage{natbib}
\usepackage{graphicx}
\usepackage{amsmath}

\begin{document}

\title{Coexistence of orbital and quantum critical magnetoresistance in FeSe$_{1-x}$S$_{x}$}

\author{S. Licciardello}
\affiliation{High Field Magnet Laboratory (HFML-EMFL) and Institute for Molecules and Materials, Radboud University, Toernooiveld 7, 6525 ED Nijmegen, Netherlands}

\author{N. Maksimovic}
\affiliation{Department of Physics, University of California and Materials Science Division, Lawrence Berkeley National Laboratory, Berkeley, California 94720, USA}

\author{J. Ayres}
\affiliation{High Field Magnet Laboratory (HFML-EMFL) and Institute for Molecules and Materials, Radboud University, Toernooiveld 7, 6525 ED Nijmegen, Netherlands}
\affiliation{H. H. Wills Physics Laboratory, University of Bristol, Tyndall Avenue, BS8 1TL, United Kingdom}

\author{J. Buhot}
\affiliation{High Field Magnet Laboratory (HFML-EMFL) and Institute for Molecules and Materials, Radboud University, Toernooiveld 7, 6525 ED Nijmegen, Netherlands}

\author{M. \v{C}ulo}
\affiliation{High Field Magnet Laboratory (HFML-EMFL) and Institute for Molecules and Materials, Radboud University, Toernooiveld 7, 6525 ED Nijmegen, Netherlands}

\author{B. Bryant}
\affiliation{High Field Magnet Laboratory (HFML-EMFL) and Institute for Molecules and Materials, Radboud University, Toernooiveld 7, 6525 ED Nijmegen, Netherlands}

\author{S. Kasahara}
\affiliation{Department of Physics, Kyoto University, Sakyo-ku, Kyoto 606-8502, Japan}

\author{Y. Matsuda}
\affiliation{Department of Physics, Kyoto University, Sakyo-ku, Kyoto 606-8502, Japan}

\author{T. Shibauchi}
\affiliation{Department of Advanced Materials Science, University of Tokyo, Kashiwa, Chiba 277-8561, Japan}

\author{V. Nagarajan}
\affiliation{Department of Physics, University of California and Materials Science Division, Lawrence Berkeley National Laboratory, Berkeley, California 94720, USA}

\author{J. G. Analytis}
\affiliation{Department of Physics, University of California and Materials Science Division, Lawrence Berkeley National Laboratory, Berkeley, California 94720, USA}

\author{N. E. Hussey}
\email[]{nigel.hussey@ru.nl}
\affiliation{High Field Magnet Laboratory (HFML-EMFL) and Institute for Molecules and Materials, Radboud University, Toernooiveld 7, 6525 ED Nijmegen, Netherlands}

\date{\today}

\begin{abstract}
The recent discovery of a non-magnetic nematic quantum critical point (QCP) in the iron chalcogenide family FeSe$_{1-x}$S$_{x}$ has raised the prospect of investigating, in isolation, the role of nematicity on the electronic properties of correlated metals. Here we report a detailed study of the normal state transverse magnetoresistance (MR) in FeSe$_{1-x}$S$_{x}$ for a series of S concentrations spanning the nematic QCP. For all temperatures and \textit{x}-values studied, the MR can be decomposed into two distinct components: one that varies quadratically in magnetic field strength $\mu_{0}\textit{H}$ and one that follows precisely the quadrature scaling form recently reported in metals at or close to a QCP and characterized by a \textit{H}-linear MR over an extended field range. The two components evolve systematically with both temperature and S-substitution in a manner that is determined by their proximity to the nematic QCP. This study thus reveals unambiguously the coexistence of two independent charge sectors in a quantum critical system. Moreover, the quantum critical component of the MR is found to be less sensitive to disorder than the quadratic (orbital) MR, suggesting that detection of the latter in previous MR studies of metals near a QCP may have been obscured.
\end{abstract}

\maketitle

\section{introduction}
Many strongly interacting electron systems lie in close proximity to a QCP, realized by suppressing a finite temperature ordering transition to zero temperature via some non-thermal tuning parameter \cite{Sachdev2011}. Metallic quantum critical systems exhibit anomalous transport and thermodynamic properties, including (but not restricted to) a \textit{T}-linear resistivity at low temperatures \cite{Lohneysen1998,Custers2003,Bruin2013} and a logarithmic divergence of the electronic specific heat \cite{Lohneysen1996}. Recently, a new feature of metallic quantum criticality was discovered in the transverse magnetoresistance (whereby the magnetic field is applied perpendicular to the current) in the iron pnictide compound BaFe$_2$(As$_{1-x}$P$_x$)$_2$ (Ba122) near its antiferromagnetic QCP \cite{Hayes2016}. In particular, the magneto-resistivity, when expressed as $\Delta\rho/T$ (= $\rho[H,T]-\rho[0,0]$) was found to exhibit an unusual quadrature scaling form $\sqrt{1+\gamma(\mu_B\mu_0H/k_BT)^2}$ where $0.5 \leq \gamma \leq 1$ is a dimensional parameter, $k_B$ is Boltzmann’s constant and $\mu_B$ is the Bohr magneton \cite{Hayes2016, Hayes2018}. Thus, in addition to a \textit{T}-linear resistivity at zero-field, $\Delta\rho$ is found to vary linearly with magnetic field strength over a wide field range. A similar scaling of the transverse MR was also reported recently in the electron-doped cuprate La$_{2-x}$Ce$_x$CuO$_4$ (LCCO), again near its antiferromagnetic QCP \cite{Sarkar2018}.

In ordinary metals, the low-field orbital MR $\delta\rho/\rho[0,T] = (\rho[H,T]-\rho[0,T])/\rho[0,T] \propto (\omega_c\tau)^2$ where $\omega_c = e\mu_0H/m^*$ is the cyclotron frequency, $m^*$ is the effective mass of the charge carriers, \textit{e} is the electric charge and $\tau$ the scattering time \cite{footnote}. In the limit where $\omega_c\tau < 1$, $\delta\rho/\rho[0,T]$ thus varies quadratically with field and given that $\rho[0,T] \propto 1/\tau$, the transverse MR has a strong temperature dependence that often obeys another form of scaling, known as Kohler's scaling, in which plots of $\delta\rho/\rho[0,T]$ versus $(H/\rho[0,T])^2$ at different temperatures collapse onto a single curve \cite{Pippard1989}. In certain correlated metals, such as the hole-doped cuprates \cite{Harris1995} and the heavy fermion CeCoIn$_5$ \cite{Nakajima2007}, a modified Kohler's scaling is observed, whereby plots of $\delta\rho/\rho[0,T]$ versus $(H/\tan{\Theta_H})^2$ collapse onto a single curve, where $\tan{\Theta_H}$ is the tangent of the Hall angle. By contrast, the MR curves in Ba122 and LCCO display no intrinsic temperature dependence -- they simply present a set of parallel curves (at high field) offset by the change in $\rho[0,T]$ \cite{Hayes2016,Hayes2018,Sarkar2018,Giraldo-Gallo2018}. 

At present, there is no consensus as to the origin of the quadrature form for the transverse MR in QC metals nor for the violation of Kohler scaling in other highly correlated metals. Moreover, it is not known how these very distinct MR responses are related, if at all. In particular, there is, as yet, no system in which signatures of the different MR behavior have been shown to co-exist, suggesting that they are forms associated with different limits (e.g. the low– and high-field limits or the behavior of systems located near or far from a QCP). 

In this contribution, we report the observation of two essentially additive components in the transverse MR of a series of FeSe$_{1-x}$S$_{x}$ single crystals that collectively span a QCP, in this case a nematic QCP. One component has a quadratic-in-field MR response up to the highest fields studied (in all S-doped samples), suggesting that this $H^2$ MR is not the limiting low-field form of the quadrature component, but something distinct, presumably reflecting the (near-)perfect compensation of the electron and hole carriers in this family of semi-metals. The second component, obtained by subtracting the $H^2$ term, exhibits the quadrature scaling form to a very high degree of precision, unambiguously demonstrating its coexistence with the conventional, orbital contribution. The two components evolve systematically with both temperature and S-substitution in a manner that is determined by the proximity to the QCP. This study thus reveals the coexistence of two charge sectors in a quantum critical system. Moreover, comparison of the MR response of two samples with very different residual resistivities reveal a marked difference in the sensitivity of the two components to disorder.  

\section{Nematic quantum criticality in the iron chalcogenides}

\begin{figure}
	\includegraphics[width=0.45\textwidth]{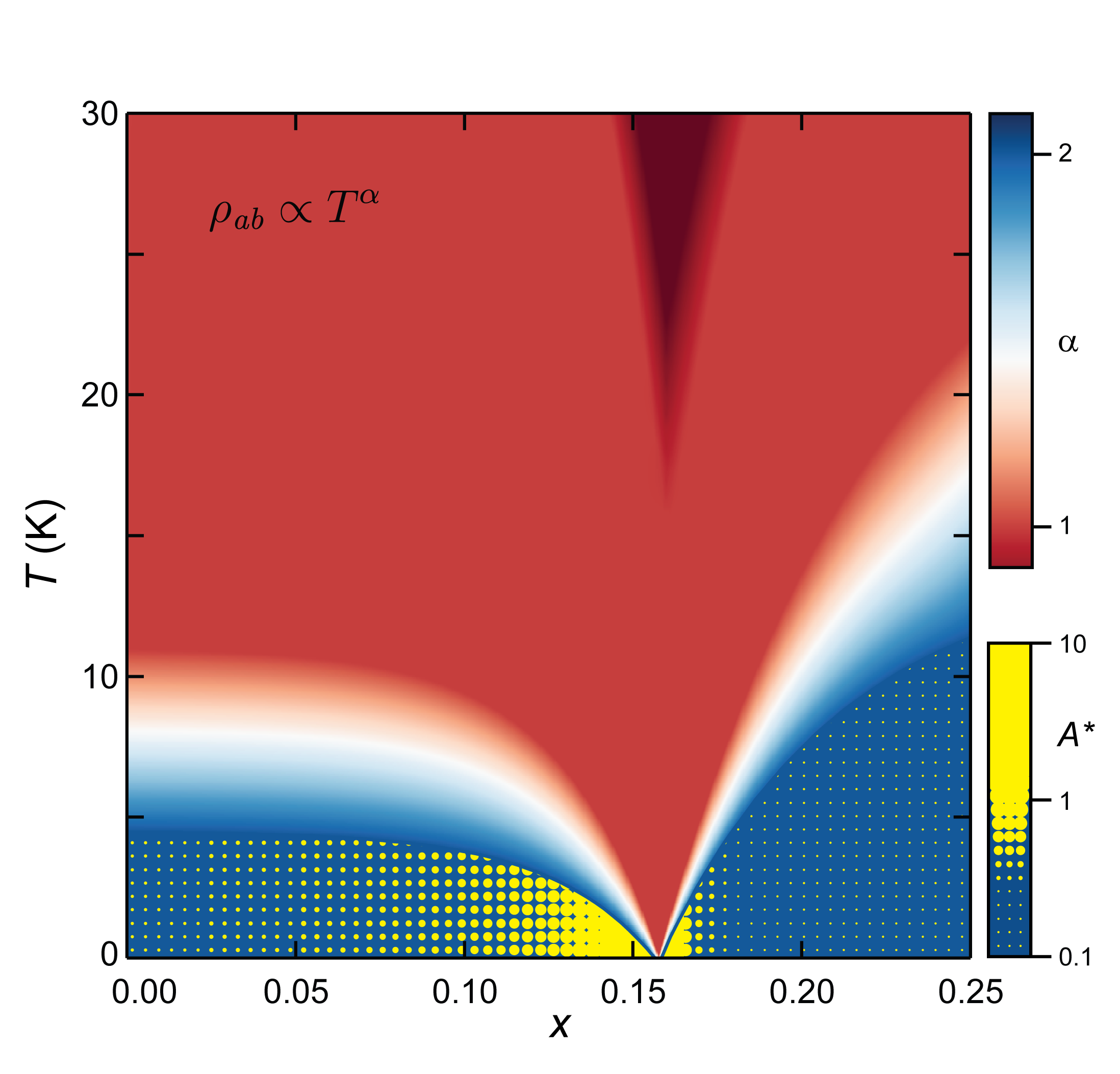}
	\caption{\label{recap} Low-temperature phase diagram of FeSe$_{1-x}$S$_{x}$ described in terms of the exponent of the \textit{T}-dependent resistivity that is itself defined in the upper color scale. The size of dots inside the $T^2$ regime indicate the strength of $A^*$, the coefficient of the $T^2$ resistivity, normalized to a fixed carrier density \cite{Licciardello2019}.}
\end{figure}

The iron chalcogenide family FeSe$_{1-x}$S$_{x}$ represents a class of quantum critical metals in which the QCP is due to electronic nematicity rather than antiferromagnetism \cite{Hosoi2016,Baek2015,Watson2015a,Watson2015b}. Recently, the evolution of the (in-plane) resistivity across the nematic QCP was studied in high magnetic fields applied in the longitudinal field configuration (\textbf{H}//\textit{I}//\textit{ab}) in order to suppress superconductivity while at the same time, minimizing the normal state MR \cite{Licciardello2019}. To orientate the subsequent analysis and discussion, we reproduce in Fig. \ref{recap} a schematic of the low-temperature phase diagram of FeSe$_{1-x}$S$_{x}$ as deduced from the temperature-dependent exponent $\alpha$ of the in-plane resistivity across the doping series at temperatures below 30 K \cite{Licciardello2019}. The top color scale in Fig. \ref{recap} denotes the magnitude of $\alpha$ at different \textit{T} and \textit{x}. At \textit{x} = \textit{x}$_c$, $\rho[T]$ is \textit{T}-linear down to 1.5 K while on either side of the QCP, $\rho[T]$ was found to crossover  to a $T^2$ dependence characteristic of a correlated Fermi liquid. $A^*$ -- the coefficient of the $T^2$ resistivity (once corrected for the growth in total carrier density with S-doping) -- was found to become strongly enhanced on approach to \textit{x}$_c$ (from either side), as indicated by the lower color scale. All these observations are consistent with those found in other quantum critical metals and suggest a strong coupling of the charge carriers to quantum fluctuations of the relevant order parameter.

It should be acknowledged here that there is currently no recognized theory for a \textit{T}-linear resistivity down to \textit{T} = 0 at a nematic QCP in a clean system \cite{Wang2019}. While FeSe exhibits only nematic order below $T_s$, a spin density wave (SDW) state is found to be stabilized under applied pressure \cite{Bendele2010}. Moreover, enhanced spin fluctuations (at ambient pressure) and critical behavior have been reported below $T_s$ \cite{Wiecki2017,Grinenko2018}, in the same range over which $\rho_{ab}[T]$ is quasi-\textit{T}-linear, suggesting a possible link between the \textit{T}-linear resistivity and antiferromagnetic, rather than nematic fluctuations. With increasing S substitution, however, the nematic and SDW states become decoupled \cite{Matsuura2017}, and as the pressure range of nematic order shrinks, eventually vanishing at \textit{x}$_c$, the dome of SDW orders shifts to progressively higher pressures. Thus, at \textit{x} = \textit{x}$_c$, the SDW phase is located far from the ambient pressure axis at which our experiments are performed. At the same time, nuclear magnetic resonance (NMR) experiments have shown that spin fluctuations, although present in FeSe$_{1-x}$S$_{x}$ at low \textit{x} values, are strongly suppressed with S substitution \cite{Wiecki2018}. These combined results suggest that the critical behavior at \textit{x} = \textit{x}$_c$ cannot be associated with proximity to a magnetic phase.

\begin{figure*}
	\includegraphics[width=0.95\textwidth]{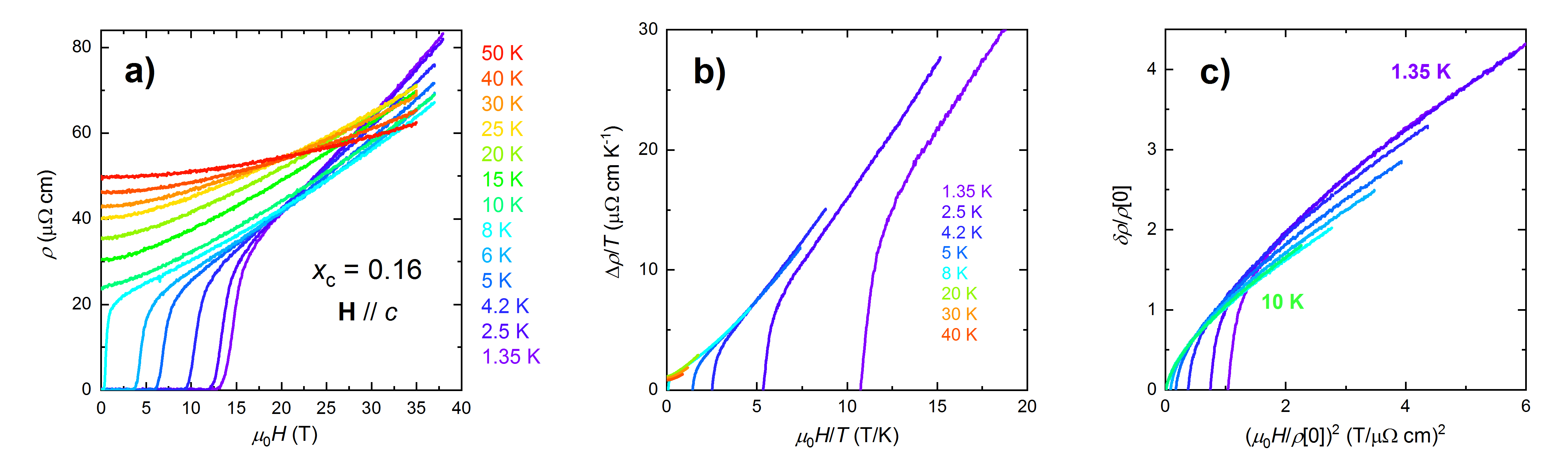}
	\caption{\label{MR016} Set of transverse MR curves for FeSe$_{0.84}$S$_{0.16}$ (i.e. at \textit{x} = \textit{x}$_c$) up to 38 T for 1.35 K $\leq$ \textit{T} $\leq$ 50 K (individual \textit{T} labels are given in the panels). Note the numerous crossing points -- behavior distinct from that found recently in other QC systems \cite{Hayes2016,Hayes2018,Sarkar2018}. b) Test for QC scaling in FeSe$_{0.84}$S$_{0.16}$. Plot of $\Delta\rho/T$ versus $\mu_0H/T$ over the same temperature range, where $\Delta\rho/T = \rho[H,T] - \rho[0,0]$. c)  Test for Kohler's scaling in FeSe$_{0.84}$S$_{0.16}$. Plot of $\delta\rho/\rho[0]$ versus $(\mu_0H/\rho[0])^2$ where $\delta\rho = \rho[H,T] - \rho[0]$ and $\rho[0]$ = $\rho[0,T]$.}
\end{figure*}

\section{methods}

The single crystals of FeSe$_{1-x}$S$_{x}$ used in this study were grown at two different locations. The bulk of the samples were grown in Kyoto by the chemical vapor transport technique \cite{Hosoi2016}. The actual sulfur composition \textit{x} was determined by the energy dispersive X-ray (EDX) spectroscopy, and was found to be around 80\% of the nominal S content. The Berkeley sample discussed exclusively in section \ref{discussion} was grown using the KCl flux technique \cite{Ma2014} with a nominal concentration of 18\% selenium replaced by sulfur, whose composition was also confirmed by EDX. To be consistent with the data presented in Ref. \cite{Licciardello2019} (carried out on the same Kyoto crystals), all \textit{x} values quoted here are the nominal values. The crystals were cut into regularly-shaped platelets and electrical contacts applied to each sample in a Hall bar geometry.  The magnetotransport measurements were carried out at the High Field Magnet Laboratory (HFML) in Nijmegen in a resistive Bitter magnet with a maximum field of 38 T using a combination of He-4 and He-3 cryostats and at the National High Magnetic Field Laboratory (NHMFL) in Los Alamos in a pulsed magnet with a field strength of 60 T. For the HFML experiments, the orientation of the samples with respect to the applied magnetic field was determined first by using a Hall probe to orient the rotating platform, then the MR of the sample itself in order to locate the transverse field orientation more precisely.

\section{results and analysis}
\subsection{Transverse magnetoresistance}

Fig. \ref{MR016} shows a series of transverse MR (\textbf{H}//\textit{c}) curves between 1.35 K and 50 K for a FeSe$_{1-x}$S$_{x}$ single crystal at the QCP ($x_c$ = 0.16) whose in-plane resistivity was found to be \textit{T}-linear down to the lowest temperatures studied. In contrast to other quantum critical systems (i.e. Ba122 and LCCO), where the MR curves taken at different temperatures are found to be simply shifted vertically with respect to one another, $\delta\rho[\mu_0H]$ in FeSe$_{0.84}$S$_{0.16}$ is found to show a strong \textit{T}-dependence with multiple crossings. Consequently, when plotted as $\Delta\rho/T$ versus $\mu_0H/T$ (Fig. \ref{MR016}b) the MR sweeps do not fall onto a single curve, except in a narrow temperature range 4.2 K $\leq$ \textit{T} $\leq$ 20 K. Even in this intermediate range, however, the form of the MR does not follow the quadrature scaling ansatz. Moreover, as shown in Fig. \ref{MR016}c, Kohler's scaling is not obeyed either. The same is true for the entire series of Kyoto samples that have been investigated.

\begin{figure*}
	\includegraphics[width=0.95\textwidth]{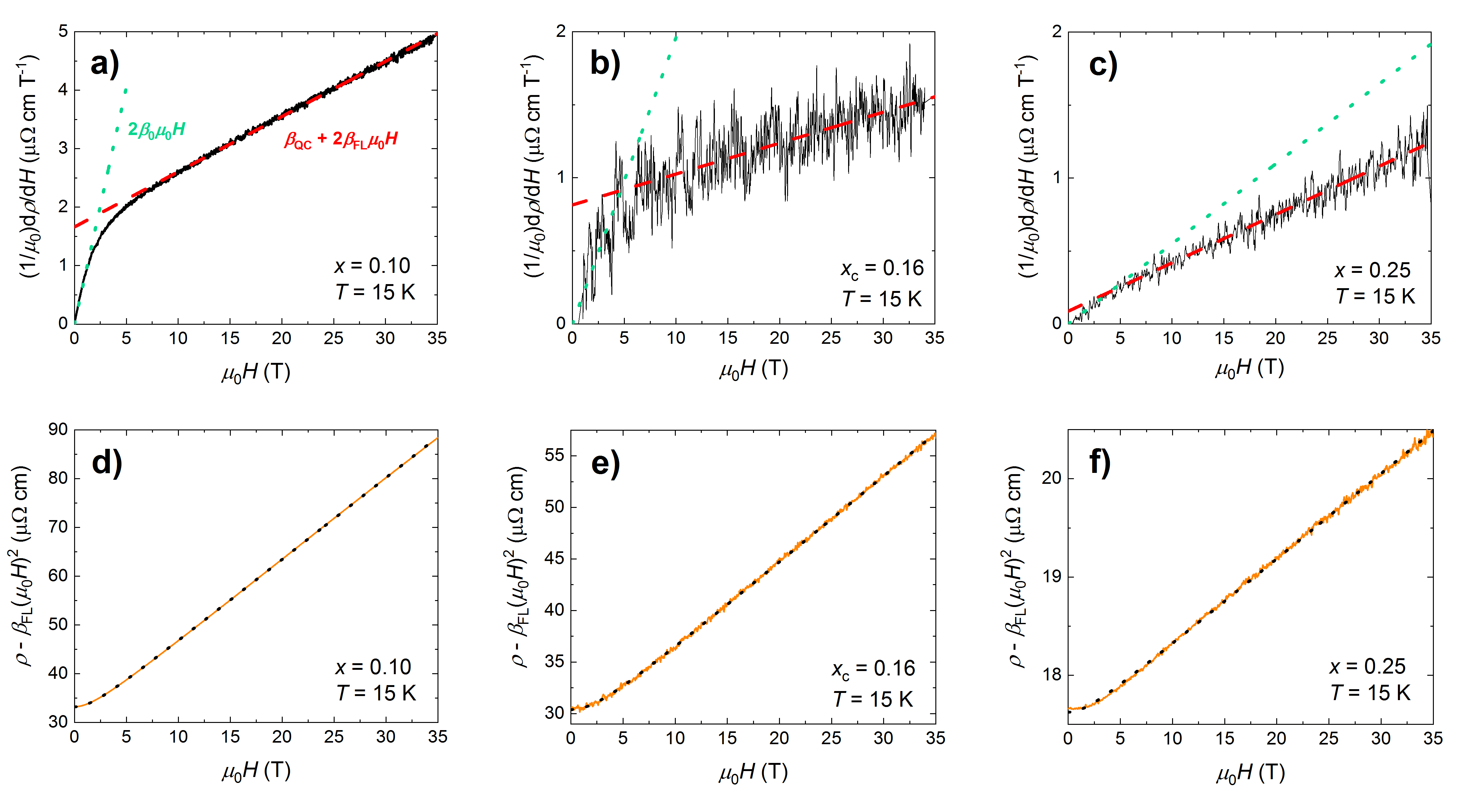}
	\caption{\label{QCcomp} d$\rho$/d($\mu_0H$) versus $\mu_0H$ at \textit{T} = 15 K for a) \textit{x} = 0.10, b) $x_c$ = 0.16 and c) \textit{x} = 0.25. The dotted lines indicate the low-field $H^2$ dependence while the dashed lines highlight the high-field $H + H^2$ dependence. The magnitude of each component is labelled $\beta_0$, $\beta_{\rm{QC}}$ and $\beta_{\rm{FL}}$ as defined in the text. Upon subtraction of the high-field $H^2$ component, one obtains the corresponding \lq residual' MR terms shown in panels d)-f) for \textit{x} = 0.10, 0.16 and 0.25 respectively. The dashed lines are fits to the quadrature form $a \sqrt{1+b(\mu_0H)^2}$.}
\end{figure*}

The reason for this lack of scaling in either $\delta\rho/\rho[0,T]$ or $\Delta\rho/T$ becomes apparent when one inspects the derivative d$\rho$/d($\mu_0H$) of the individual MR curves. Panels a) – c) in Figure \ref{QCcomp} show d$\rho$/d($\mu_0H$) curves for \textit{x} = 0.10, 0.16 and 0.25 respectively obtained at \textit{T} = 15 K, where superconducting fluctuations are effectively suppressed. While the specific form of the derivative is most evident in the \textit{x} = 0.10 sample (Fig. 3a), qualitatively similar behavior is found for all the other samples. 

At the lowest fields, d$\rho$/d($\mu_0H$) is linear in field with a zero intercept, implying that the low-field MR is strictly quadratic. The slope of the derivative is labeled 2$\beta_0\mu_0H$ and is indicated in each case by a green dotted line. For 2 T $< \mu_0H <$ 7 T, the slope of d$\rho$/d($\mu_0H$) gradually decreases until above 7 T, it becomes linear once more, albeit with a finite intercept. The presence of this finite intercept implies that for $\mu_0H >$ 7 T, the MR has two components, one linear in field, the other quadratic. Both components persist up to the highest field measured. Note that such a field dependence cannot be captured by a simple two-carrier model involving electrons and holes \cite{Rourke2010}.

The slope of d$\rho$/d($\mu_0H$) at high field is defined here as 2$\beta_{\rm{FL}}$ where $\beta_{\rm{FL}}$ is the magnitude of the $H^2$ term that we argue below arises from orbital (i.e. cyclotron) effects. Upon subtracting this term from the total MR, the form of the second component in $\delta\rho[\mu_0H]$ is revealed. As indicated by the dotted black lines in panels d) – f), the remaining contribution to the MR is found to follow the same quadrature form, i.e. $\rho[\mu_0H]$ - $\beta_{FL}(\mu_0H)^2$ = $a \sqrt{1+b(\mu_0H)^2}$ that was first reported in BaFe$_2$(As$_{1-x}$P$_x$)$_2$ \cite{Hayes2016} (here \textit{a} and \textit{b} are fitting parameters). The quality of the fit, over the entire field range studied, appears to confirm that the transverse MR of FeSe$_{0.9}$S$_{0.1}$ comprises two distinct terms, one that is quadratic at all fields, and one that possesses the quadrature form (note that $\beta_0$ is a compound term, comprising both $\beta_{\rm{FL}}$ and the low-field quadratic part of the quadrature MR).

Further derivatives and residual MR curves for different samples recorded at different temperatures are presented in Fig. \ref{QCcomp-suppl} of the Supplemental Material. Significantly, the same features are observed for all \textit{x} and \textit{T}, albeit with different relative weightings, implying that these two distinct MR contributions persist over the entire range of temperatures and S concentrations studied. Only in stoichiometric FeSe, where the orbital MR is extremely large, is this term found to deviate from $H^2$ at high fields and low \textit{T}, though even here, the form of the MR (having subtracted off the quadrature component) is consistent with the usual Drude expression for two-carrier (i.e. electron and hole) magnetotransport (see Fig. \ref{anomalx0-suppl} in Supplemental Material for more detail). 

\subsection{Two component magnetoresistance}

The data presented in Figure \ref{QCcomp} reveal a novel aspect of the MR response in FeSe$_{1-x}$S$_x$, namely the presence of two contributions which individually extend over a wide field, temperature and doping range. In light of this, we conclude that the charge dynamics of FeSe$_{1-x}$S$_x$ must contain two distinct sectors, one that generates a conventional orbital MR, presumably involving quasiparticle transport, and one akin to the quantum critical sector found in Ba122 and LCCO (that exhibits scale-invariance). In such a scenario, the total (zero-field) conductivity $\sigma_{\rm{tot}}$ should be expressed as a sum of the individual contributions, i.e. $\sigma_{\rm{tot}}$[\textit{T}] = $\sigma_{\rm{QC}}$[\textit{T}] + $\sigma_{\rm{FL}}$[\textit{T}], where the subscripts refer to the quantum critical and quasiparticle (Fermi-liquid) sectors respectively. The transverse magneto-conductance is then given by a weighted sum \cite{Hussey1998}:

\begin{equation}
\label{twocompMC}
\frac{\Delta\sigma_{\rm{tot}}}{\sigma_{\rm{tot}}}=\frac{\sigma_{\rm{QC}}}{\sigma_{\rm{tot}}}\frac{\Delta\sigma_{\rm{QC}}}{\sigma_{\rm{QC}}}+\frac{\sigma_{\rm{FL}}}{\sigma_{\rm{tot}}}\frac{\Delta\sigma_{\rm{FL}}}{\sigma_{\rm{FL}}}
\end{equation}

In reality, of course, it is the magnetoresistance, rather than the magneto-conductance that is measured, the former being related to the latter via inversion of the (in-plane) conductivity tensor.

\begin{equation}
\label{MR}
\frac{\delta\rho}{\rho[0,T]}=-\frac{\Delta\sigma_{\rm{tot}}}{\sigma_{\rm{tot}}} - \bigg(\frac{\sigma_{\rm{xy}}}{\sigma_{\rm{tot}}}\bigg)^2
\end{equation}

where $\rho[0,T] = 1/$$\sigma_{\rm{tot}}$ is the zero-field resistivity at the temperature at which an individual field sweep is taken, $\sigma_{\rm{xy}}$ is the Hall conductivity and $\sigma_{\rm{xy}}$/$\sigma_{\rm{tot}}$ the corresponding Hall angle. In order to proceed, it is necessary to estimate first the magnitude of the Hall angle ($\sigma_{\rm{xy}}$/$\sigma_{\rm{tot}}$) relative to $\delta\rho/\rho[0,T]$ (as measured). In the temperature range 20 K $< T <$ 50 K over which we currently have data overlap, the square of the Hall angle (for \textit{x} = 0.17) is found to be $\sim$ 20 \% of the measured MR with only a small (25 \%) variation in \textit{T} \cite{Hosoiprv}. By contrast, the MR changes by a factor of four (\textit{x} = 0.16) or five (\textit{x} = 0.18) across the same \textit{T} range. Thus, we can conclude that the MR is dominated by the magneto-conductance term and re-write Eq. \ref{twocompMC} as:

\begin{equation}
\label{twocompMR}
\frac{\delta\rho}{\rho[0,T]}=\frac{\sigma_{\rm{QC}}}{\sigma_{\rm{tot}}}\frac{\delta\rho_{\rm{QC}}}{\rho_{\rm{QC}}[0,T]}+\frac{\sigma_{\rm{FL}}}{\sigma_{\rm{tot}}}\frac{\delta\rho_{\rm{FL}}}{\rho_{\rm{FL}}[0,T]}
\end{equation}

Hence,

\begin{equation}
\label{twocomprho1}
\delta\rho[H]=\bigg(\frac{\sigma_{\rm{QC}}}{\sigma_{\rm{tot}}}\bigg)^2\delta\rho_{\rm{QC}}[H]+\bigg(\frac{\sigma_{\rm{FL}}}{\sigma_{\rm{tot}}}\bigg)^2\delta\rho_{\rm{FL}}[H]
\end{equation}

which at high fields can be expressed as:

\begin{equation}
\label{twocomprho2}
\delta\rho[H]=\beta_{\rm{QC}}\mu_0H+\beta_{\rm{FL}}(\mu_0H)^2 
\end{equation}

Here $\beta_{\rm{QC}}$ and $\beta_{\rm{FL}}$ are, respectively, the (as-measured) magnitudes of the \textit{H}-linear and $H^2$ MR terms, which according to Eq. \ref{twocomprho1} and Eq. \ref{twocomprho2} represent the quantum critical $\delta\rho_{\rm{QC}}[H]$ and quasiparticle $\delta\rho_{\rm{FL}}[H]$ contributions to the total MR, weighted by the square of the contribution of the two sectors to the total (zero-field) conductivity.

\subsection{Evolution across the phase diagram}

\begin{figure}
	\includegraphics[width=0.5\textwidth]{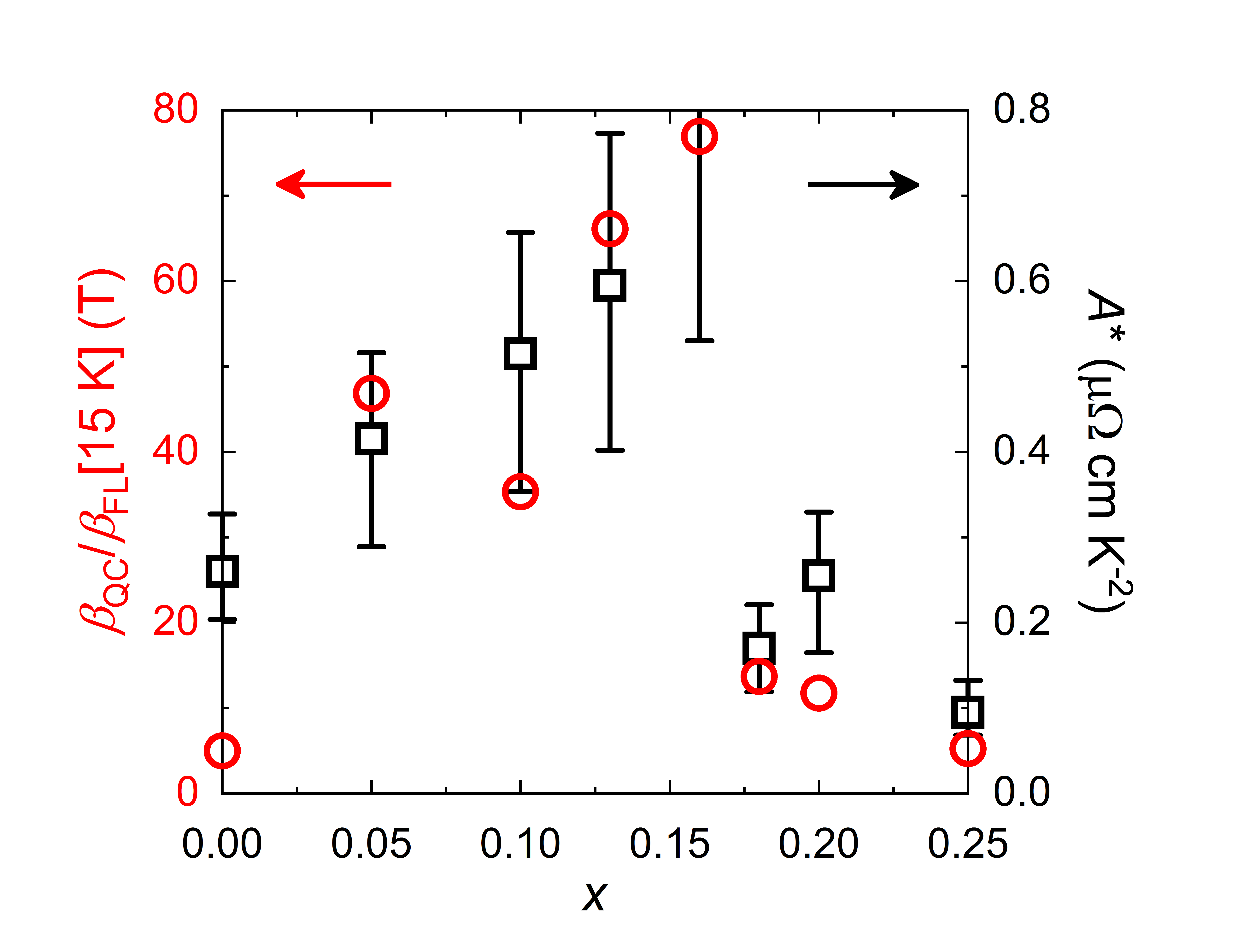}
	\caption{\label{ratios} (Open circles) Variation of $\beta_{\rm{QC}}$/$\beta_{\rm{FL}}$, the ratio of the \textit{H}-linear transverse MR to the (high-field) $H^2 $ component as a function of \textit{x}. All $\beta_{\rm{QC}}$/$\beta_{\rm{QC}}$ values were obtained at \textit{T} = 15 K. (Open squares) Corresponding values of $A^*$, the coefficient of the $T^2$ resistivity, normalized to a fixed carrier density \cite{Licciardello2019}.}
\end{figure}

The ratio $\beta_{\rm{QC}}$/$\beta_{\rm{FL}}$ for all samples - determined at a temperature (15 K) at which there are no discernible superconducting fluctuation conductivity -  is plotted in Figure \ref{ratios} (as open red circles). The ratio is found to peak around $x_c$ = 0.16, in a manner that is strikingly similar to the enhancement of the quasiparticle effective mass as expressed through $A^*$, the renormalized coefficient of the $T^2$ resistivity (and plotted as empty black squares in Fig. \ref{ratios}) \cite{Licciardello2019}. It is important to realize that these two quantities are determined in very different ways yet together, they appear to confirm the influence of quantum critical fluctuations on the electrical resistivity of FeSe$_{1-x}$S$_x$.

As described above, plots of $\Delta\rho/T$ versus $H/T$ in P-doped Ba122 at the critical doping are found to collapse onto a single curve of the quadrature form \cite{Hayes2016}. Such scaling can only be realized if the high-field (\textit{H}-linear) slopes of the individual MR curves are the same, i.e. $\Delta\rho_{\rm{QC}}$ = $X_1\mu_0H$ independent of temperature. Since there are two contributions to the MR in FeSe$_{1-x}$S$_x$, whose relative strengths are weighted by their respective contributions to the total conductivity, the same $\Delta\rho/T$ scaling cannot be gleaned directly from our data by simply subtracting off the orbital MR term. Nevertheless, further analysis outlined below and presented in sections \ref{scaling15SM} and \ref{scaling010SM} of the Supplemental Material provides strong evidence that $H/T$ scaling is also realized in FeSe$_{1-x}$S$_x$.

Firstly, according to the scaling ansatz of Hayes \textit{et al.} \cite{Hayes2016}, the residual MR (obtained by subtracting the $H^2$ term from the total MR) should have the same dependence with field for all samples when measured at the same temperature, irrespective of its absolute magnitude. As shown in Fig. \ref{scaling15-suppl} of the Supplemental Material, the (normalized) residual MR at \textit{T} = 15 K is indeed found to follow the same form right across the phase diagram. Secondly, when the residual MR for one sample is plotted versus $H/T$ for a range of temperatures inside the QC fan (see Fig. \ref{scaling010-suppl} of Supplemental Material), the data are found to collapse onto a single curve.  Finally, as described in the Discussion section, a second sample with a doping close to $x_c$ but with a larger residual resistivity (that effectively quenches the orbital component to the MR), is found to exhibit precisely the same MR scaling as seen in Ba122 and LCCO. Thus, we can conclude that the QC component to the MR in FeSe$_{1-x}$S$_x$ follows the exact same scaling relation, and since d[$\Delta\rho_{\rm{QC}}$]/d\textit{H} = d[$\delta\rho_{\rm{QC}}$]/d\textit{H} (only the intercepts differ), we arrive at the following relation between $\beta_{\rm{QC}}$ and $X_1$

\begin{equation}
\label{betaqc1}
\beta_{\rm{QC}}=\bigg(\frac{\sigma_{\rm{QC}}}{\sigma_{\rm{tot}}}\bigg)^2X_1
\end{equation}

\begin{figure*}
	\includegraphics[width=0.95\textwidth]{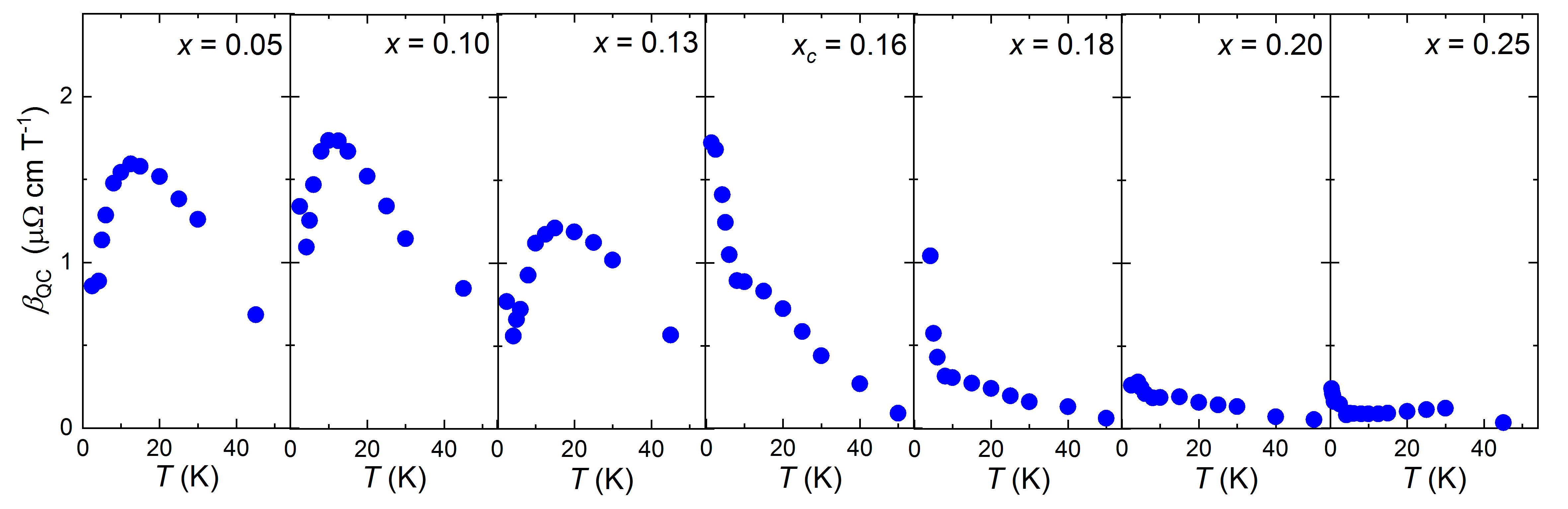}
	\caption{\label{betaQC} Temperature and \textit{x} dependence of $\beta_{\rm{QC}}$, the strength of the \textit{H}-linear transverse magnetoresistance in FeSe$_{1-x}$S$_x$ for all the Kyoto samples.}
\end{figure*}

Thus, under the inference that the QC component to the MR in FeSe$_{1-x}$S$_x$ exhibits scale invariance, $\beta_{\rm{QC}}$ provides a direct measure of the contribution of $\sigma_{\rm{QC}}$, the QC component, to the total conductivity. This quantity is plotted in Figure \ref{betaQC} for all the S concentrations studied (bar \textit{x} = 0.00 for which $\beta_{\rm{QC}}$ is hard to extract due to its exceptional high-field behavior). What is most striking here is the evolution in the behaviour of $\beta_{\rm{QC}}$[\textit{T}] across $x_c$. For samples with $x < x_c$, $\beta_{\rm{QC}}$[\textit{T}] follows the same \textit{T}-dependence, reaching a maximum at or around the temperature below which $\rho(T)$ is no-longer \textit{T}-linear, i.e. below the QC fan, implying that the QC component is reduced as one approaches the FL ($\rho \sim T^2$) regime (see Fig. \ref{recap}). By contrast, for $x_c$ = 0.16, $\beta_{\rm{QC}}$ increases monotonically with decreasing temperature, consistent with the observation that the \textit{T}-linear resistivity extends down to the lowest \textit{T} accessed to date and indicating that as the temperature is lowered, the QC component emerges as the dominant contribution. Finally, beyond  $x_c$, the magnitude of $\beta_{\rm{QC}}$ is much reduced, though in contrast to the samples in  the nematic phase, it appears to show an upturn at the lowest temperatures, whose magnitude gradually softens with further S doping. Indeed, for \textit{x} = 0.25, $\beta_{\rm{QC}}$ has almost disappeared. 

\begin{figure}
	\includegraphics[width=0.4\textwidth]{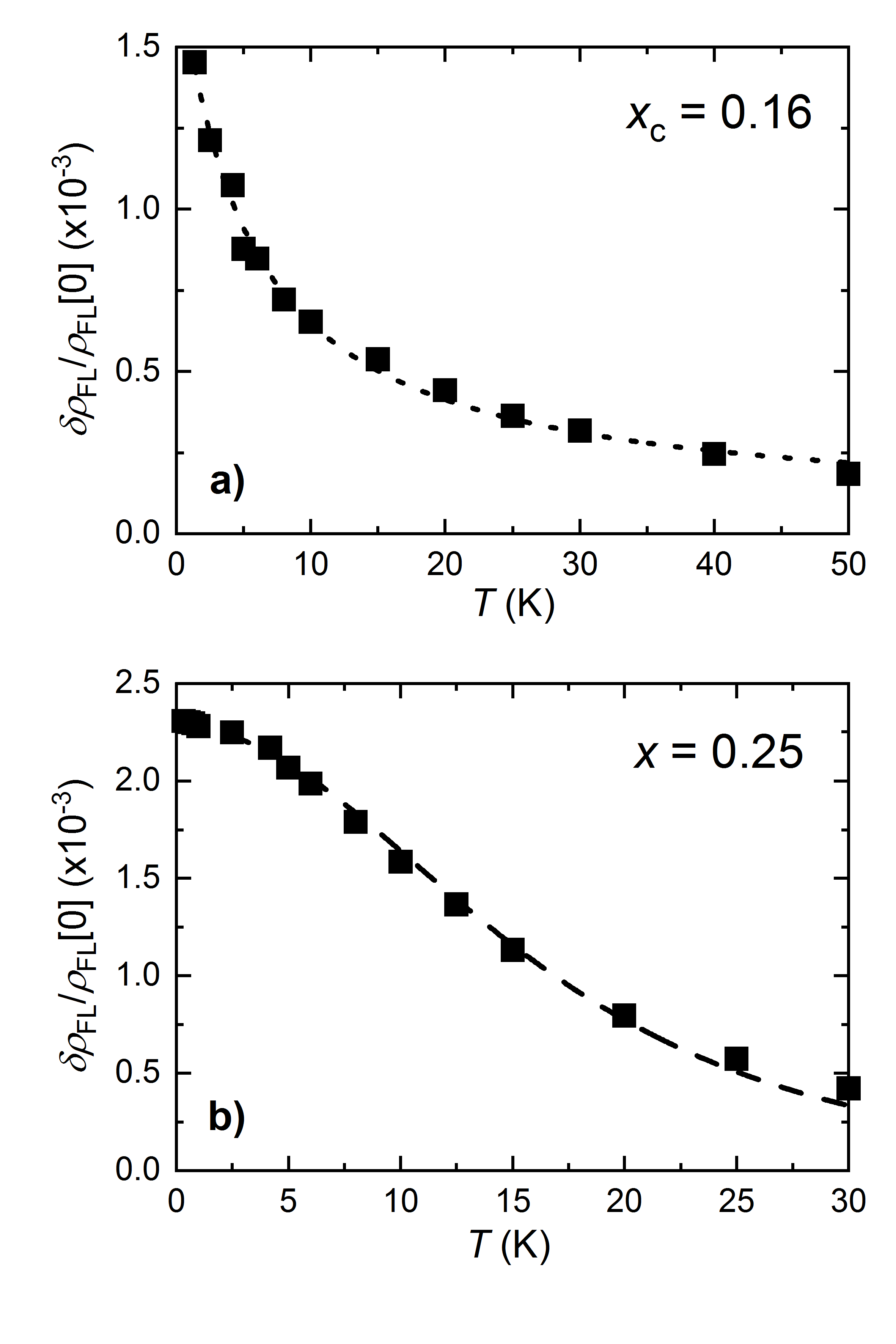}
	\caption{\label{FLcomp} \textit{T}-dependence of $\delta\rho_{\rm{FL}}/\rho_{\rm{FL}}[0]$ (at $\mu_0H$ = 1 T) for a) $x_c$ = 0.16 and b) \textit{x} = 0.25. The dotted line in panel a) is a guide to the eye. The dashed line in panel b) denotes a fit to the data to the expression $\delta\rho_{\rm{FL}}/\rho_{\rm{FL}}[0]$ = 1/($A + BT^2$)$^2$ up to 30 K, as expected in a correlated Fermi liquid.}
\end{figure}

Of course, while $\beta_{\rm{QC}}$ is proportional to ($\sigma_{\rm{QC}}$/$\sigma_{\rm{tot}})^2$, we cannot determine $\sigma_{\rm{QC}}$/$\sigma_{\rm{tot}}$ directly as we have no way of obtaining $X_1$ independently. However, one can gain an estimate for $\sigma_{\rm{QC}}$/$\sigma_{\rm{tot}}$ by simulating the zero-field $\rho(T)$ assuming parallel conduction, a point we shall return to later. Nevertheless, Fig. \ref{betaQC} reveals a very systematic evolution in the fraction of the total conductivity that can be attributed to the QC component.

\begin{figure}
	\includegraphics[width=0.4\textwidth]{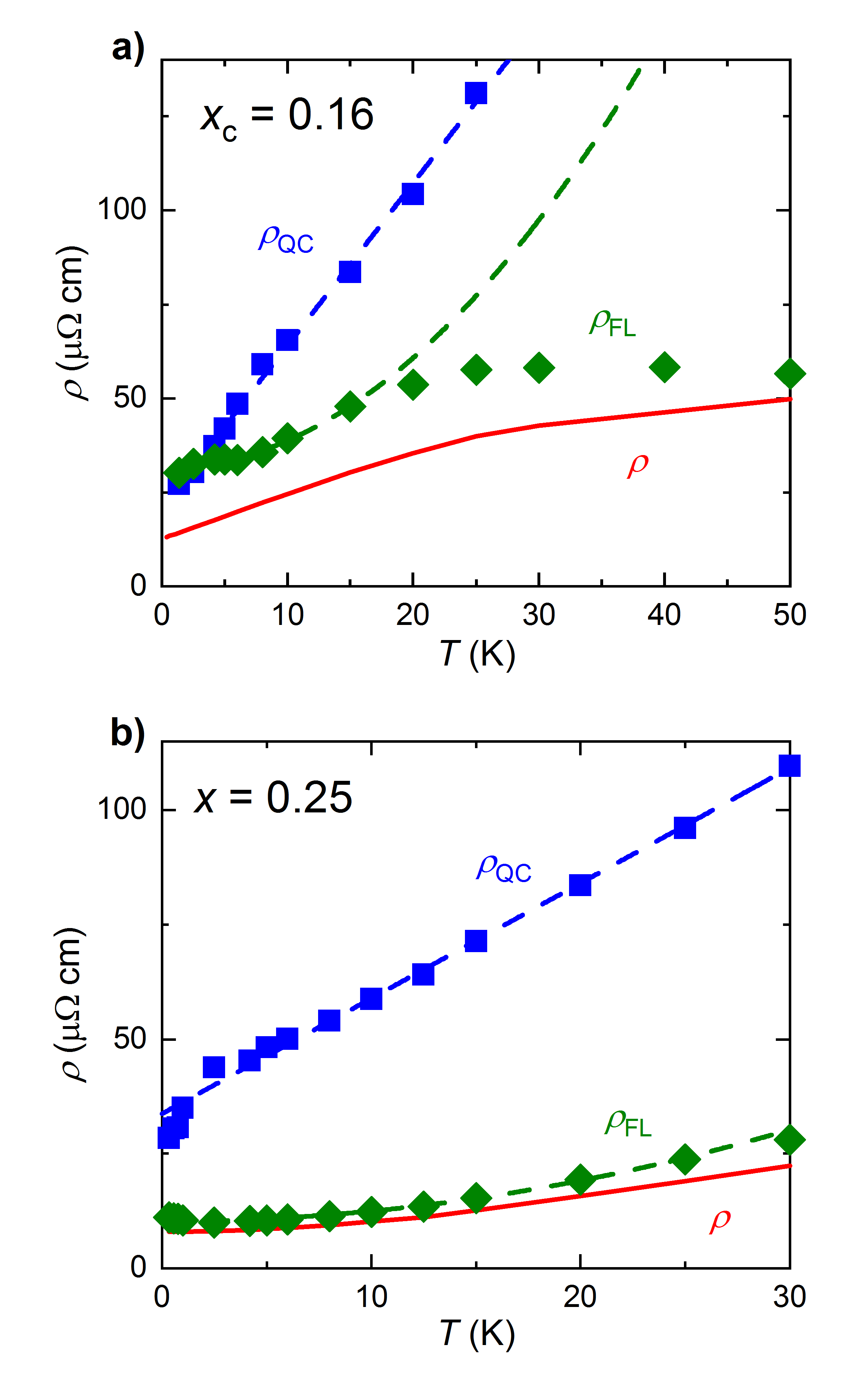}
	\caption{\label{deconvolution} Decomposition of experimentally determined $\rho(T)$ (solid lines) for a) FeSe$_{0.84}$S$_{0.16}$ and b) FeSe$_{0.75}$S$_{0.25}$ into quantum critical and quasiparticle channels obtained from the transverse MR study. The blue and green dashed lines represent \textit{T}-linear and $T^2$ dependencies respectively. In both cases, 1/$\rho$ = 1/$\rho_{\rm{FL}}$ + 1/$\rho_{\rm{QC}}$.}
\end{figure}

For completeness, we now turn to consider the second component $\delta\rho_{\rm{FL}}=\beta_{\rm{FL}}(\mu_0H)^2$. The large field range over which this MR component remains strictly quadratic, suggests that the electron and hole pockets in our FeSe$_{1-x}$S$_x$ crystals are close to being fully compensated. Moreover, as shown in Figure \ref{FLcomp}b for the \textit{x} = 0.25 sample, the \textit{T}-dependence of $\delta\rho_{\rm{FL}}/\rho_{\rm{FL}}[0]$ (where $\rho_{\rm{FL}}[0]$ = $\rho_{\rm{FL}}[0,T]$ is estimated from the zero-field resistivity shown in Fig. \ref{deconvolution} and discussed below) is found to have a Fermi-liquid (FL) form; $\delta\rho_{\rm{FL}}/\rho_{\rm{FL}}[0]$ = 1/($\omega_c\tau$)$^2$ = 1/($A + BT^2$)$^2$ between 1 K and 30 K. At the QCP however, $\delta\rho_{\rm{FL}}/\rho_{\rm{FL}}[0]$ does not follow the same form (Fig. \ref{deconvolution}a). Nevertheless, the very strong \textit{T}-dependence observed in both cases supports the notion that this contribution is controlled by orbital effects (i.e. by $\omega_c\tau$).

\subsection{Two component conductivity}

The presence of two distinct components in the transverse MR of FeSe$_{1-x}$S$_x$ implies that there must also be two contributions to the zero-field conductivity, i.e. $\sigma_{\rm{tot}}$ = $\sigma_{\rm{FL}}$ + $\sigma_{\rm{QC}}$; the first term giving rise to the conventional, orbital MR and the second to the QC quadrature term. While the QC component $\sigma_{\rm{QC}}$ is linked directly to $\beta_{\rm{QC}}$ through Eq. \ref{twocomprho1} and \ref{twocomprho2}, it cannot be determined uniquely since $X_1$ itself is not known. We can, however, allow the magnitude of $X_1$ to vary and inspect the resultant \textit{T}-dependence of $\rho_{\rm{QC}}$ = 1/$\sigma_{\rm{QC}}$ and $\rho_{\rm{FL}}$ = 1/$\sigma_{\rm{FL}}$ = 1/($\sigma_{\rm{tot}}$ - $\sigma_{\rm{QC}}$) where $\sigma_{\rm{tot}}$ = 1/$\rho$[0,\textit{T}], in order to see whether or not a self-consistent picture for both the zero-field resistivity and the transverse MR emerges from the data.

Examples of this procedure are shown in Figures \ref{deconvolution}a) and \ref{deconvolution}b) for the $x_c$ = 0.16 and \textit{x} = 0.25 samples with $X_1$ = 2.5 and 1.7 respectively. Here, we have ensured that the two components add in parallel to give the total, as measured, resistivity. $\rho_{\rm{QC}}(T)$ is found to be \textit{T}-linear in both cases, at least up to 25 K. For \textit{x} = 0.25, $\rho_{\rm{FL}}(T)$ retains its $T^2$ character up to 30 K, even though the raw resistivity curve is quadratic only up to 12 K. For $x_c$ = 0.16, it is difficult to distinguish between a quadratic or linear form for $\rho_{\rm{FL}}(T)$ below 15 K, and as shown in Ref. \cite{Licciardello2019}, the normal-state resistivity (having suppressed the superconductivity in a longitudinal magnetic field) appears to be strictly \textit{T}-linear down to 1.5 K. However, the data presented in Fig. \ref{deconvolution}a) are not inconsistent with the coexistence both the QC and the FL components in the resistivity, while the presence of the quasiparticle component in the transverse MR may indicate that this sample is located very close to, though not necessarily at the QCP. Further measurements down to lower temperatures (in higher fields) will be needed to determine the form of $\beta_{\rm{QC}}$/$\beta_{\rm{FL}}$ at $x = x_c$ below 1.5 K.

\section{\label{discussion}discussion}

The observation of two distinct components in the transverse MR of FeSe$_{1-x}$S$_x$ raises the question why previous studies of correlated metals (both in the vicinity of or far from a QCP) found only an orbital MR response (that may or may not have violated Kohler's rule) or the quadrature scaling form, but never the combination. A comparative study of two crystals with different levels of disorder, presented below, provides one possible explanation for these distinct behaviors.

\begin{figure*}
	\includegraphics[width=0.95\textwidth]{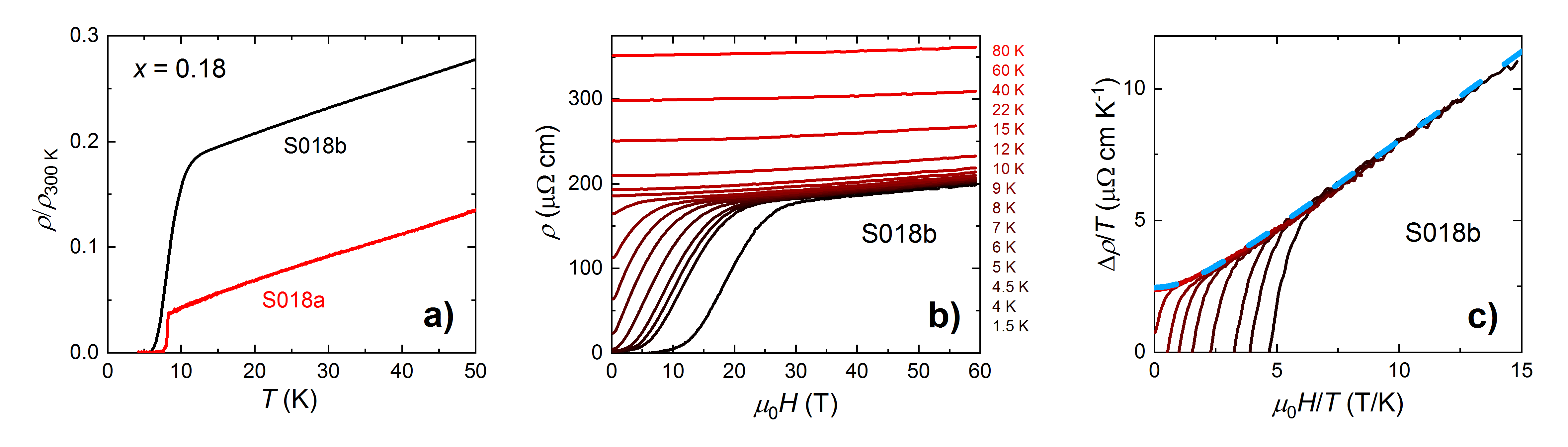}
	\caption{\label{disorder} a) Zero-field resistivity for two FeSe$_{0.82}$S$_{0.18}$ crystals grown via different techniques (see text for details).  b) Set of transverse MR curves for S018b up to 60 T obtained at temperatures as labelled. (c) Scaling plot of $\Delta\rho/T$ vs. $\mu_0H/T$ for S018b. The dashed line is the quadrature fit of the form $\sqrt{1+\gamma(\mu_B\mu_0H/k_BT)^2}$ (with $\gamma \approx$ 0.5).}
\end{figure*}

Figure \ref{disorder}a) shows the low-\textit{T} resistivity of the two crystals in question (both with nominal composition \textit{x} = 0.18). The crystal with the lower residual resistivity (S018a) was synthesized in Kyoto using identical starting constituents and growth conditions as the other crystals described in the preceding section. As with the other crystals from this source, it exhibits both components in the transverse MR that evolve with temperature as summarized in Figures \ref{ratios} and \ref{betaQC}. The second crystal (S018b) was prepared in Berkeley using a different technique and found to have a residual resistivity that is approximately 5 times higher. A series of MR curves obtained on this crystal over a wide temperature range (1.5 K $< T <$ 80 K) is shown in Fig. \ref{disorder}b). In contrast to the multiple crossing points realized in the other crystals (an example of which is shown in Fig. \ref{MR016}a)), the MR curves for S018b (beyond the field-induced superconductor-to-metal transition) are parallel to one another and become \textit{H}-linear at high fields. Moreover, when the MR curves are re-plotted as $\Delta\rho/T$ versus $H/T$, as shown in Fig. \ref{disorder}c), they are found to collapse onto a single curve that fits the same quadrature form $\Delta\rho/T = \sqrt{1+\gamma(\mu_B\mu_0H/k_BT)^2}$  (with $\gamma \approx$ 0.5) that was observed in the cleaner crystal (though now unfettered by the presence of the orbital MR term).

The observation of QC scaling in the MR response of the second crystal reveals that while the orbital component is effectively quenched with increasing impurity scattering (a five-fold increase in the residual resistivity would correspond to a 25-fold decrease in the orbital MR at low-\textit{T}), the QC component survives. 

It has been argued previously that the \textit{H}-linear transverse MR and $H/T$ scaling found in pnictides \cite{Hayes2016} and cuprates \cite{Sarkar2018} may arise due to a variation in the carrier composition \cite{Singleton2018}, as postulated previously for two-dimensional electron gases \cite{Khouri2016} and even elemental metals \cite{Bruls1981}. The current study, however, suggests that this is not necessarily the case in QC systems. As shown in Fig. \ref{ratios}, the ratio of the \textit{H}-linear ($\beta_{\rm{QC}}$) to $H^2$ ($\beta_{\rm{FL}}$) components shows a very systematic evolution with S substitution and peaks strongly at the QCP, even though the residual resistivities are comparable across the entire series of S-doped crystals \cite{Licciardello2019}. Moreover, $\beta_{\rm{QC}}$/$\beta_{\rm{FL}}$ in FeSe is the same as in FeSe$_{0.75}$S$_{0.25}$, despite the fact that the former's residual resistivity is one order of magnitude smaller than the latter and clear quantum oscillations are observed in the former. Finally, the sharpness of the kinks in d$\rho$/d\textit{T} \cite{Licciardello2019} at \textit{T} = $T_s$ (for $x  < x_c$) imply homogeneous doping for all these samples. Thus it appears unlikely that the \textit{H}-linear component to the transverse MR is due to an extrinsic longitudinal contribution arising from a variation in carrier density along each crystal.

Recent models of strange metals, invoking either holographic methods \cite{Cremonini2017} or based on the Sachdev-Ye-Kitaev picture of itinerant but non-quasiparticle transport \cite{Patel2018}, have succeeded in obtaining certain aspects of the MR scaling, but as of yet, not in tandem with a more conventional, orbital MR. The key task now therefore is to understand how can these two components can coexist. 

We consider here first the possibility that the two components arise from excitations that occupy different regions of the Brillouin zone. FeSe and its derivatives are known to contain (equal) numbers of electron- and hole-like carriers and correlated metals often display an electron-hole dichotomy, most evident in the respective phase diagrams of electron- and hole-doped cuprates, for example \cite{Armitage2010}. In such a scenario, the electron and hole pockets found in FeSe$_{1-x}$S$_x$ would harbour different types of excitations that contribute respectively to the orbital and QC MR responses. Alternatively, the two excitations may reside within both pockets, albeit at different points on the Fermi surface; for example, the QC component may arise from excitations near hot-spots -- strong scattering sinks that destroy the quasiparticle character of excitations there -- leading to strong momentum dependent scattering as realized, e.g. in the cuprates \cite{Hussey2018}. Indeed, the superconducting gap in FeSe has been shown to be strongly anisotropic in both the electron and hole pockets, indicating anisotropic (and possibly orbitally-selective) pairing interactions \cite{Sprau2017}. Both scenarios, however, appear inconsistent with the observation of quantum oscillations on both the electron and hole pocket (at least for S concentrations located away from the QCP) that indicate the presence of coherent quasiparticle states around the Fermi surface of both pockets \cite{Terashima2014,Coldea2019}. 

The lack of oscillations at $x = x_c$ itself, however, is consistent with the persistence of the non-FL resistivity down to the lowest temperatures at this concentration and it is there that the QC component of the MR is most dominant. The gradual crossover from quantum critical to quasiparticle contributions to the MR away from the QCP suggests in fact that the low-lying excitations near the Fermi level have dual character, i.e. the quasiparticle and the quantum critical sectors are two \lq flip sides' of the same electronic states, whose weighting depends on their proximity to the QCP. Whatever the origin, these findings clearly call for further theoretical studies in order to understand the interplay of the two sectors across the phase diagram, and more experimental studies to determine quantitatively the role of disorder in the realization of the $H/T$ scaling in the transverse MR not only in more disordered FeSe$_{1-x}$S$_x$, but also in other candidate QC systems. The latter comparison is important to establish whether it is merely a question of disorder or length scales (e.g. between electron-electron collisions and electron-impurity collisions), or whether it is the nematic character of the quantum fluctuations in FeSe$_{1-x}$S$_x$ that allows both the quantum critical and quasiparticle sectors to reveal themselves, even at the QCP itself.

\section{conclusions}

In summary, we have carried out a systematic study of the transverse magnetoresistance (MR) in a series of FeSe$_{1-x}$S$_x$ single crystals in high magnetic fields up to 38 T for S concentrations that span the nematic quantum critical point (QCP). The field derivatives of the MR curves reveal the ubiquitous presence of two distinct (and additive) components to the MR in FeSe$_{1-x}$S$_x$; the normal orbital $H^2$ MR and an anomalous component that follows precisely the quadrature scaling first observed in the iron pnictide P-doped BaFe$_2$As$_2$ near the spin-density-wave QCP.

The ratio of the two MR components follows a very similar evolution with doping as the (renormalized) \textit{A} coefficient of the $T^2$ resistivity, indicating that the component with the quadrature form is also associated with scale-invariant quantum critical fluctuations that are also responsible for the quasiparticle mass enhancement on approaching the QCP. The quantum critical contribution is found to become enhanced with decreasing \textit{T} at the QCP, but is suppressed inside the FL regime away from the QCP. With increased disorder content, the orbital MR is quenched, leading to the appearance of strict quantum critical scaling at or near to the QCP.

These collective findings provide the first evidence for the coexistence of two charge sectors in a quantum critical system whose relative weighting evolves systematically with proximity to the QCP. The task now is to identify how these two sectors co-exist and to establish whether this is a universal behavior in quantum critical systems that, until now, may have been obscured by the presence of disorder.

\begin{acknowledgments}
The authors would like to thank M. Bristow, A. Coldea, B. Gouteraux, M. Katsnelson, A. Krikun, P. Reiss, K. Schalm and J. Schmalian for enlightening discussions on the work presented here. We also acknowledge the support of the HFML-RU/NWO, a member of the European Magnetic Field Laboratory (EMFL), and the NHMFL Pulsed Field Facility at Los Alamos, New Mexico. This work is part of the research programme ‘Strange Metals’ (grant number 16METL01) of the former Foundation for Fundamental Research on Matter (FOM), which is financially supported by the Netherlands Organisation for Scientific Research (NWO). A portion of this work was also supported by the Engineering and Physical Sciences Research Council (grant number EP/L015544/1) and by Grants-in-Aid for Scientific Research (KAKENHI) (grant numbers 15H02106, 15H03688, 15KK0160, 18H01177 and 18H05227) and Innovative Areas ‘Topological Material Science’ (grant number 15H05852) from the Japan Society for the Promotion of Science (JSPS). VN and NM were supported by the Gordon and Betty Moore Foundation’s EPiQS Initiative through Grant GBMF4374. JGA acknowledges partial support from the Center for Novel Pathways to Quantum Coherence in Materials, an Energy Frontier Research Center funded by the U.S. Department of Energy, Office of Science, Basic Energy Sciences.
.
\end{acknowledgments}

\clearpage
\newpage

\pagebreak
\widetext
\begin{center}
	\textbf{\large Supplemental Material}
\end{center}

\setcounter{equation}{0}
\setcounter{figure}{0}
\setcounter{table}{0}
\setcounter{page}{1}
\setcounter{section}{0}
\makeatletter
\renewcommand{\theequation}{S\arabic{equation}}
\renewcommand{\thefigure}{S\arabic{figure}}

\section{Deconvolution of the transverse MR in F\lowercase{e}S\lowercase{e$_{1-x}$}S$_x$}

\begin{figure*}[h]
	\includegraphics[width=0.9\textwidth]{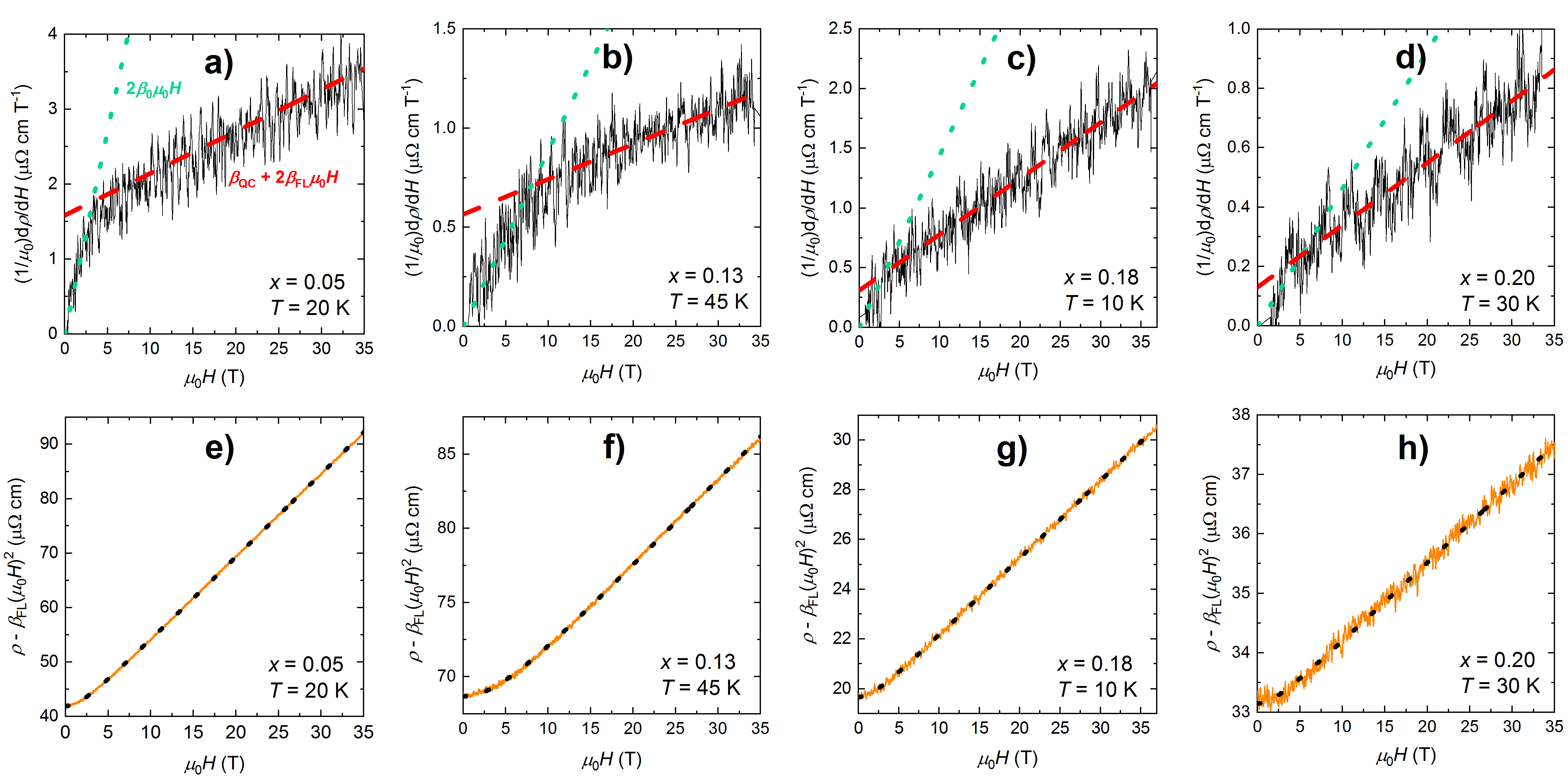}
	\caption{\label{QCcomp-suppl}Panels a)-d) d$\rho$/d($\mu_0H$) versus $\mu_0H$ obtained from MR sweeps performed on the other S-doped single crystals at different temperatures as indicated in the individual panels. In all cases, upon subtraction of the strictly $H^2$ orbital component (indicated by a thick dashed line), one obtains the corresponding \lq residual' MR terms shown in panels e)-h) that follow the quadrature scaling form $a \sqrt{1+b(\mu_0H)^2}$ (shown as a black dotted line) to a high degree of precision. The green dotted lines in panels a)-d) indicate the low-field $H^2$ dependence that forms part of the residual (quadrature) MR, as explained in the main text.}
\end{figure*}

\section{Deconvolution of the transverse MR in undoped F\lowercase{e}S\lowercase{e}}

\begin{figure*}
	\includegraphics[width=0.9\textwidth]{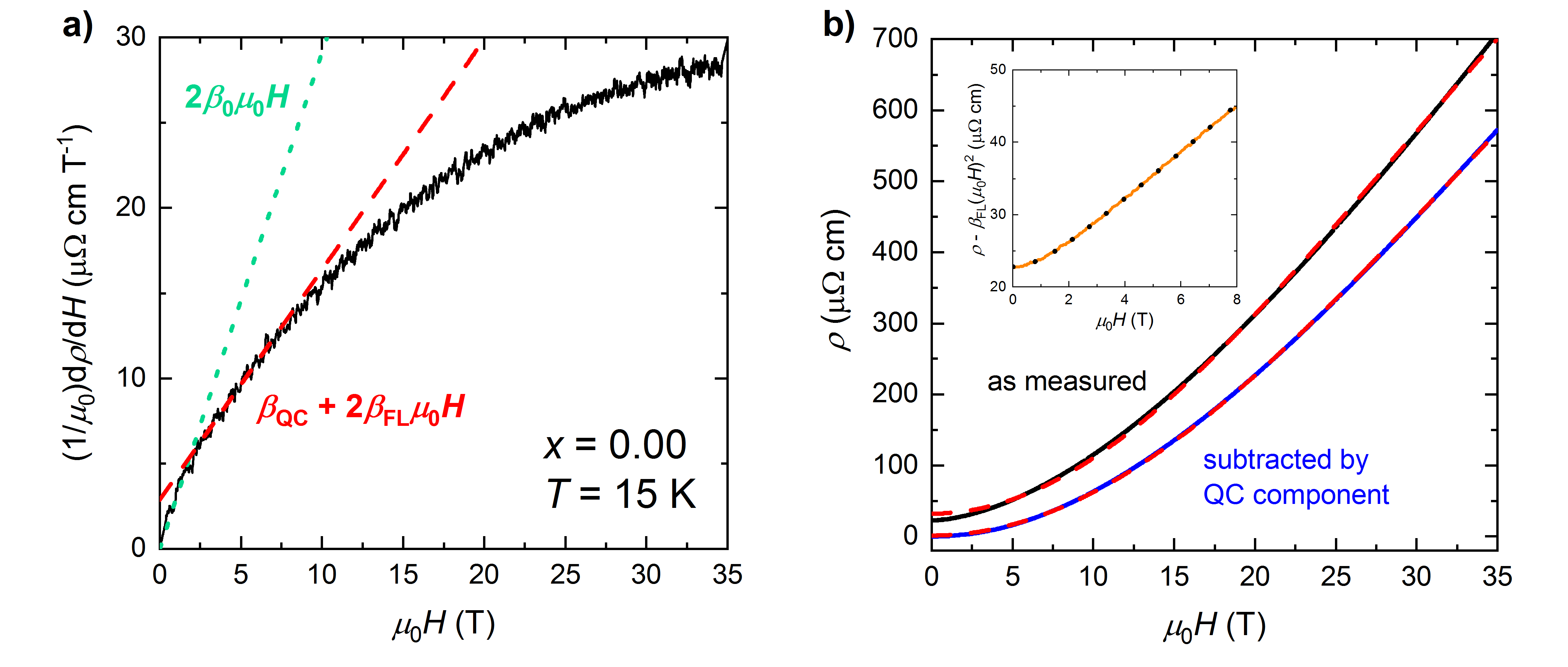}
	\caption{\label{anomalx0-suppl} a) d$\rho$/d($\mu_0H$) versus $\mu_0H$ for undoped FeSe at \textit{T} = 15 K. b) Inset: Upon subtraction of the $\beta_{\rm{FL}}(\mu_0H)^2$ orbital component below 7 T (indicated by a thick red dashed line) in panel a), one obtains the corresponding \lq residual' MR term that again follows the quadrature scaling form $a \sqrt{1+b(\mu_0H)^2}$ (shown as a black dotted line). Main panel: Upper curve: Total transverse MR curve fitted to the two-carrier model. Lower curve; orbital MR (obtained by subtracting the QC component with the quadrature form indicated in the inset) fitted to the same two-carrier expression.}
\end{figure*}

\begin{figure}[h!]
	\includegraphics[width=0.4\textwidth]{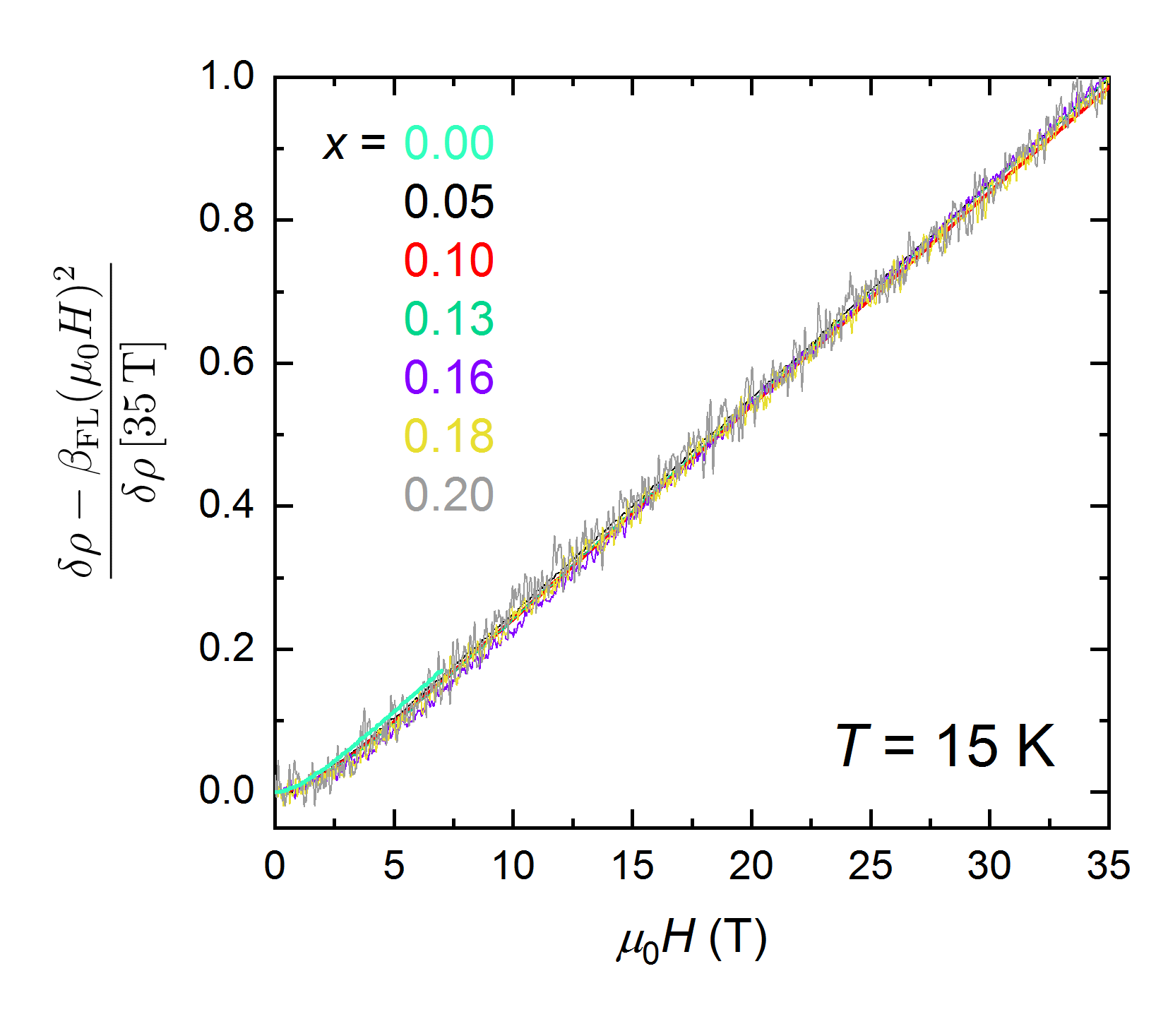}
	\caption{\label{scaling15-suppl} Residual transverse MR in FeSe$_{1-x}$S$_x$ (obtained by subtraction of $\rho(H=0)$ and the $\beta_{\rm{FL}}(\mu_0H)^2$ orbital component) across the phase diagram at \textit{T} = 15 K. All curves have been normalized to their value at 35 T.}
\end{figure}

In undoped FeSe, the transverse MR is much larger than in the S-doped crystals. As a result, the orbital component exhibits a marked deviation from the strictly $H^2$ dependence found in the other samples, even at intermediate field strengths. Nevertheless, it is still possible to fit the orbital component using the standard model for two-carrier (i.e. electron and hole) magnetoresistance. Fig. \ref{anomalx0-suppl}a) shows d$\rho$/d\textit{H} in FeSe measured at 15 K (again away from the superconducting fluctuation regime). Below 7 T, the form of the derivative is identical to that found in S-doped crystals, i.e. with an initial linear rise (reflecting a quadratic MR) followed by a crossover to a second linear regime with an intercept (corresponding to the $\beta_{\rm{QC}}\mu_0H$ + $\beta_{\rm{FL}}(\mu_0H)^2$ form of the MR seen elsewhere). Above 7 T, however, the high-field MR departs from this dependence. As shown in the inset of Fig. \ref{anomalx0-suppl}b), subtracting off the $\beta_{\rm{FL}}(\mu_0H)^2$ component from the total MR below 7 T reveals the quadrature form of the quantum critical component to the MR, which, as shown in Fig. \ref{scaling15-suppl}, follows the same form as that measured at the same temperature across the substitution series.  

The main panel in Fig. \ref{anomalx0-suppl}b) demonstrates that while the total MR in FeSe at 15 K cannot be fitted satisfactorily by the $c_1H^2$/(1 + $c_2H^2$) form characteristic of the two-carrier model \cite{Rourke2010}, upon subtraction of the QC component, the two-carrier form fits the MR perfectly. Here, we have assumed that the residual (QC) MR uncovered from analysis up to 7 T obeys the same quadrature form up to the highest field measured.

\newpage

\section{\label{scaling15SM}Test of scale invariance in the residual transverse MR in F\lowercase{e}S\lowercase{e$_{1-x}$}S$_x$ across the phase diagram}

\newpage

\section{\label{scaling010SM}Test of $H/T$ scaling in the residual transverse MR in F\lowercase{e}S\lowercase{e$_{0.9}$}S$_{0.1}$}

\begin{figure}[h!]
	\includegraphics[width=0.4\textwidth]{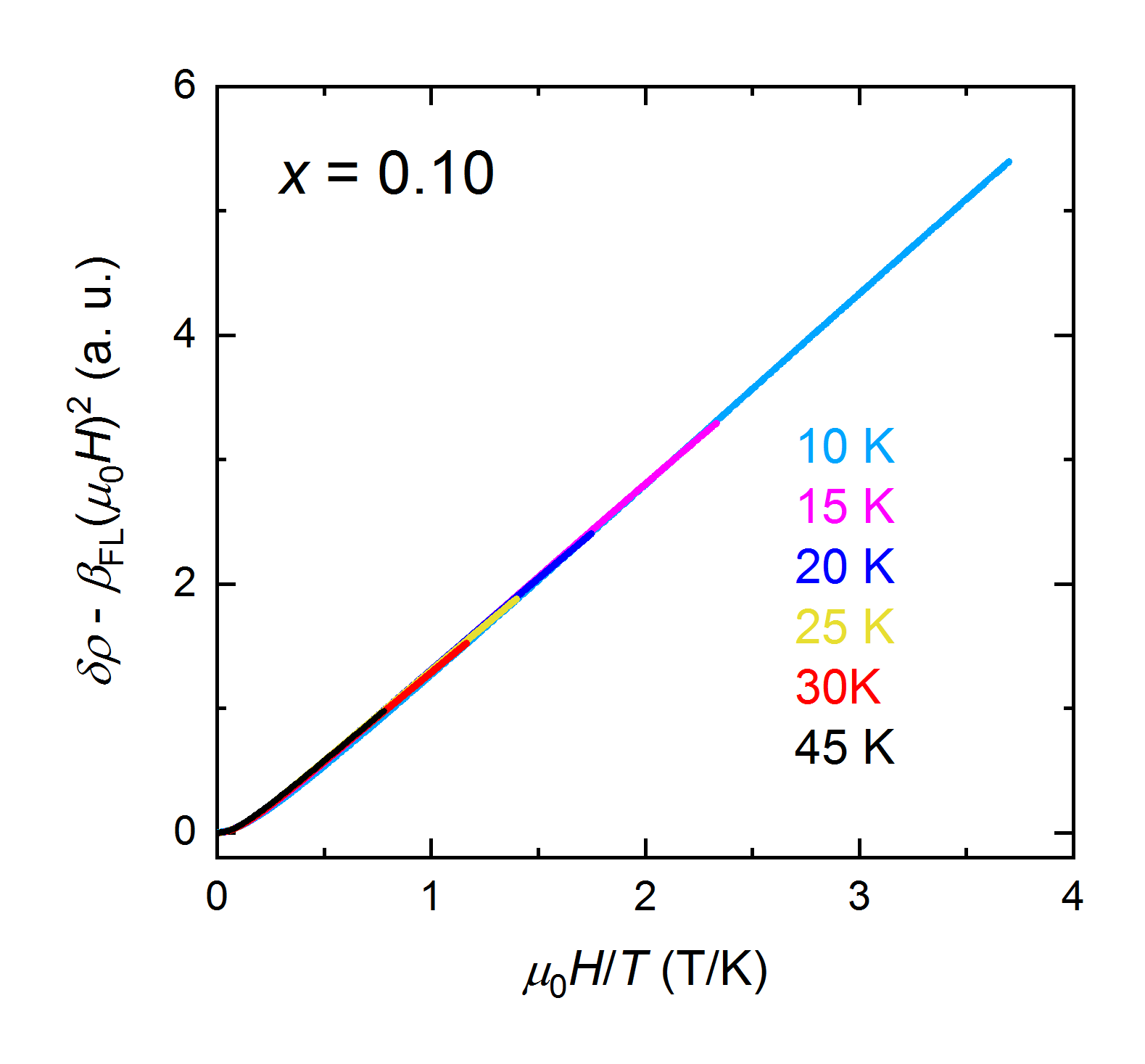}
	\caption{\label{scaling010-suppl} Renormalized residual transverse MR in FeSe$_{0.9}$S$_{0.1}$(obtained by subtraction of $\rho(H=0)$ and the $\beta_{\rm{FL}}(\mu_0)H)^2$ orbital component) at temperatures from 10 K to 45 K. Below 10 K, the low-field normal state MR is obscured by superconductivity. Hence it is not possible to test the $H/T$ scaling below this temperature.}
\end{figure}


\begin{thebibliography}{38}
	
\bibitem{Sachdev2011} S. Sachdev, \textit{Quantum Phase Transitions} (Cambridge University Press) (2011).
\bibitem{Lohneysen1998} H. v. L{\"o}hneysen, S. Mock, A. Neubert, T. Pietrus, A. Rosch, A. Schr{\"o}der, O. Stockert and U. Tutsch, J. Mag. Mag. Mater. {\bf 177-181}, 12--17 (1998).
\bibitem{Custers2003} J. Custers, P. Gegenwart, H. Wilhelm, K. Neumaier, Y. Tokiwa, O. Trovarelli, C. Geibel, F. Steglich, C. P{\'e}pin and P. Coleman, Nature (London), {\bf 424}, 524--527 (2003).
\bibitem{Armitage2010} N. P. Armitage, P. Fournier and R. L. Greene, Rev. Mod. Phys. {\bf 82}, 2421--2487 (2010)
\bibitem{Bruin2013} J. A. N. Bruin, H. Sakai, R. S. Perry and A. P. Mackenzie, A. P., Science {\bf 339}, 804--807 (2013)
\bibitem{Lohneysen1996} H. v. L{\"o}hneysen, J. Phys.: Condens. Matter {\bf 8}, 9689--9706 (1996)
\bibitem{Hayes2016} I. M. Hayes, R. D. McDonald, N. P. Breznay, T. Helm, P. J. W. Moll, M. Wartenbe, A. Shekhter and J. G. Analytis, Nat. Phys. {\bf 12}, 916--919 (2016)
\bibitem{Hayes2018} I. M. Hayes, Z. Hao, N. Maksimovic, S. K. Lewin, M. K. Chan, R. D. McDonald, B. J. Ramshaw, J. E. Moore and J. G. Analytis, Phys. Rev. Lett. {\bf 121}, 197002 (2018)
\bibitem{Sarkar2018} T. Sarkar, P. R. Mandal, N. R. Poniatowski, M. K. and Chan and R. L. Greene, R. L. arXiv:1810.03499v1 (2018)
\bibitem{footnote} Recall that while $\Delta\rho$ is obtained by subtracting the effective zero-temperature, zero-field resistivity value of $\rho$, $\delta\rho$ is obtained by subtracting the 0 T resistivity value at the temperature of each field sweep.
\bibitem{Pippard1989} A. B. Pippard, {\it Magnetoresistance in Metals} (Cambridge University Press) (1989).
\bibitem{Harris1995} J. M. Harris, Y. F. Yan, P. Matl, N. P. Ong, P. W. Anderson, T. Kimura and K. Kitazawa, K., Phys. Rev. Lett. {\bf 75},
1391--1394 (1995)
\bibitem{Nakajima2007} Y. Nakajima, H. Shishido, H. Nakai, T. Shibauchi, K. Behnia, K. Izawa, M. Hedo, Y. Uwatoko, T. Matsumoto, R. Settai, Y. Ōnuki, H. Kontani and Y. Matsuda, Y., J. Phys. Soc. Japan {\bf 76}, 024703 (2007)
\bibitem{Giraldo-Gallo2018} P. Giraldo-Gallo, J. A. Galvis, Z. Stegen, K. A. Modic, F. F. Balakirev, J. S. Betts, X. Lian, C. Moir, S. C. Riggs, J. Wu, A. T. Bollinger, X. He, I. Bo{\v{z}}ovi{\'{c}}, B. J. Ramshaw, R. D. McDonald, G. S. Boebinger and A. Shekhter, Science {\bf 361}, 479--481 (2018)
\bibitem{Hosoi2016} S. Hosoi, K. Matsuura, K. Ishida, H. Wang, Y. Mizukami, T. Watashige, S. Kasahara, Y. Matsuda and T. Shibauchi, Proc. Natl. Acad. Sci. USA {\bf 113}, 8139--8143 (2016)
\bibitem{Baek2015} S.-H. Baek, D. V. Efremov, J. M. Ok, J. S. Kim, J. van den Brink and B. B{\"u}chner, Nat. Mater. {\bf 14}, 210--214 (2015)
\bibitem{Watson2015a} M. D. Watson, T. K. Kim, A. A. Haghighirad, N. R. Davies, A. McCollam, A. Narayanan, S. F. Blake, Y. L. Chen, S. Ghannadzadeh, A. J. Schofield, M. Hoesch, C. Meingast, T. Wolf and A. I. Coldea, A. I., Phys. Rev. B {\bf 91}, 155106 (2015) \bibitem{Watson2015b} M. D. Watson, T. K. Kim, A. A. Haghighirad, S. F. Blake, N. R. Davies, M. Hoesch, T. Wolf and A. I. Coldea, Phys. Rev. B {\bf 92} 121108(R) (2015)
\bibitem{Licciardello2019} S. Licciardello, J. Buhot, J. Lu, J. Ayres, S. Kasahara, T. Shibauchi, Y. Matsuda and N. E. Hussey, Nature (London) {\bf 567}, 213--217 (2019)
\bibitem{Wang2019} X. Wang and E. Berg, arXiv:1902.04590 (2019)
\bibitem{Bendele2010} M., Bendele, A. Amato, K. Conder, M. Elender, H. Keller, H. H. Klauss, H. Luetkens, E. Pomjakushina, A. Raselli and R. Khasanov, Phys. Rev. Lett. {\bf 104}, 087003 (2010).
\bibitem{Wiecki2017}  P. Wiecki, M. Nandi, A. E. B{\"{o}}hmer, S. L. Bud'ko, P. C. Canfield and  Y. Furukawa, Phys. Rev. B {\bf 96}, 180502(R) (2017).
\bibitem{Grinenko2018}  V. Grinenko, R. Sarkar, P. Materne, S. Kamusella, A. Yamamshita, Y. Takano, Y. Sun, T. Tamegai, D. V. Efremov, S.-L. Drechsler, J.-C. Orain, T. Goko, R. Scheuermann, H. Luetkens and H.-H. Klauss, Phys. Rev. B {\bf 97}, 201102(R) (2018).
\bibitem{Matsuura2017}  K. Matsuura, Y. Mizukami, Y. Arai, Y. Sugimura, N. Maejima, A. Machida, T. Watanuki, T. Fukuda, T. Yajima, Z. Hiroi, K. Y. Yip, Y. C. Chan, Q. Niu, S. Hosoi, K. Ishida, K. Mukasa, S. Kasahara, J. G. Cheng, S. K. Goh, Y. Matsuda, Y. Uwatoko and T. Shibauchi, Nat. Commun. {\bf 8}, 1143 (2017).
\bibitem{Wiecki2018}  P. Wiecki, K. Rana, A. E. B{\"{o}}hmer, Y. Lee, S. L. Bud'ko, P. C. Canfield and  Y. Furukawa, Phys. Rev. B {\bf 98}, 020507(R) (2018).
\bibitem{Ma2014}  M. W. Ma, D. N. Yuan, Y. Wu, X. L. Dong and F. Zhou, Physica C {\bf 506}, 154--157 (2010).
\bibitem{Rourke2010}  P. M. C. Rourke, A. F. Bangura, C. Proust, J. Levallois, N. Doiron-Leyraud, D. LeBoeuf, L. Taillefer, S. Adachi, M. L. Sutherland and N. E. Hussey, {Phys. Rev. B {\bf 82}, 020514(R) (2010).
\bibitem{Hussey1998}  N. E. Hussey, A. P. Mackenzie, J. R. Cooper, Y. Maeno, S. Nishizaki and T. Fujita, Phys. Rev. B {\bf 57}, 5505--5511 (1998).
\bibitem{Hosoiprv}  S. Hosoi and S. Kasahara, unpublished.
\bibitem{Singleton2018}  J. Singleton, arXiv:1810.01998 (2018).
\bibitem{Khouri2016}  T. Khouri, U. Zeitler, C. Reichl, W. Wegscheider, N. E. Hussey, S. Wiedmann and J. C. Maan, Phys. Rev. Lett. {\bf 117}, 256601 (2016).
\bibitem{Bruls1981}	G. J. C. L. Bruls, J. Bass, A. P. van Gelder, H. van Kempen and P. Wyder, Phys. Rev. Lett. {\bf 46}, 553--555 (1981)
\bibitem{Cremonini2017} S. Cremonini, A. Hoover and L. Li, J. High Energ. Phys. {\bf 10} 133 (2017)
\bibitem{Patel2018} A. A. Patel, J. McGreevy, D. P. Arovas and S. Sachdev, Phys. Rev. X {\bf 8}, 021049 (2018)
\bibitem{Hussey2018} N. E. Hussey, J. Buhot and S. Licciardello, Rep. Prog. Phys. {\bf 81},	052501 (2018)
\bibitem{Sprau2017} P. O. Sprau, A. Kostin, A. Kreisel, A. E. B{\"{o}}hmer, V. Taufour, P. C. Canfield, S. Mukherjee, P. J. Hirschfeld, B. M. Andersen and J. C. Davis, Science {\bf 357}, 75--80 (2017)
\bibitem{Terashima2014} T. Terashima, N. Kikugawa, A. Kiswandhi, E.-S. Choi, J. S. Brooks, S. Kasahara, T. Watashige, H. Ikeda, T. Shibauchi, Y. Matsuda, T. Wolf, A. E. B{\"{o}}hmer, F. Hardy, C. Meingast, H. v. L{\"{o}}hneysen, M. T. Suzuki, R. Arita and S. Uji, Phys. Rev. B {\bf 90}, 144517--} (2014)
\bibitem{Coldea2019} A. I. Coldea, S. F. Blake, S. Kasahara, A. A. Haghighirad, M. D. Watson, W. Knafo, E.-S. Choi, A. McCollam, P. Reiss, T. Yamashita, M. Bruma, S. C. Speller, Y. Matsuda, T. Wolf, T. Shibauchi and A. J. Schofield, npj Quantum Mater. {\bf 4}, 2 (2019)

\end{thebibliography}
\end{document}